%% file: main.tex
\documentclass{article}

\usepackage{arxiv}
\usepackage{amssymb}
\usepackage{amsmath}
\usepackage{color}
\usepackage{caption}
\usepackage{subcaption}
\usepackage{multirow}
\usepackage{makecell}

\usepackage[utf8]{inputenc} 
\usepackage[T1]{fontenc}    
\usepackage{hyperref}       
\usepackage{url}            
\usepackage{booktabs}       
\usepackage{amsfonts}       
\usepackage{nicefrac}       
\usepackage{microtype}      
\usepackage{lipsum}
\usepackage{graphicx}

\title{Neural Eikonal Solver: improving accuracy of physics-informed neural networks for solving eikonal equation in case of caustics}

\author{
 Serafim Grubas \\
  Novosibirsk State University \\
  Russia, Novosibirsk, 630090 \\
  \texttt{serafimgrubas@gmail.com} \\
   \And
 Anton Duchkov \\
  Dynamic analysis in seismology \\
  Institute of Petroleum Geology and Geophysics\\
  Russia, Novosibirsk, 630090 \\
  \texttt{DuchkovAA@ipgg.sbras.ru} \\
  \And
 Georgy Loginov \\
  Dynamic analysis in seismology \\
  Institute of Petroleum Geology and Geophysics\\
  Russia, Novosibirsk, 630090 \\
  \texttt{loginovgeorgy@gmail.com} \\
}

\begin{document}
\maketitle
\begin{abstract}


The concept of physics-informed neural networks has become a useful tool for solving differential equations due to its flexibility. A few approaches are using this concept to solve the eikonal equation which describes the first-arrival traveltimes of acoustic and elastic waves in smooth heterogeneous velocity models. However, the challenge of the eikonal is exacerbated by the velocity models producing caustics, resulting in instabilities and deterioration of accuracy due to the non-smooth solution behaviour. In this paper, we revisit the problem of solving the eikonal equation using neural networks to tackle the caustic pathologies. We introduce the novel Neural Eikonal Solver (NES) for solving the isotropic eikonal equation in two formulations: the one-point problem is for a fixed source location; the two-point problem is for an arbitrary source-receiver pair. We present several techniques that provide relatively fast, stable, and accurate approximation of the eikonal in complex velocity models producing caustics: an improved factorization bounding the NES between the fastest and the slowest solutions to speed up the training; a non-symmetric loss function based on $L_1$-norm and a Hamiltonian of the eikonal to account for the errors caused by caustics; gaussian activation for a more accurate approximation of solution in caustics; a symmetrization to account for the reciprocity principle in the two-point problem. The tests on the Marmousi model showed that NES provides the relative mean absolute error of about 0.2-0.4\% from the second-order factored Fast Marching Method, and outperforms existing neural-network solvers giving 10-60 times lower errors and 2-30 times faster training. With using a GPU, the training takes 1-5 minutes, and the inference time is comparable with the Fast Marching. The one-point NES provides the most accurate solution, whereas the two-point NES provides slightly lower accuracy but gives an extremely compact representation. It can be useful in various seismic applications where massive computations of traveltimes are required (millions of source-receiver pairs): ray modeling, traveltime tomography, hypocenter localization, and Kirchhoff migration. Source code is available at \url{https://github.com/sgrubas/NES}
\end{abstract}


\input{Introduction}

\input{Theory}

\input{Testing}

\input{Discussion}

\input{Conlusions}

\section*{Acknowledgement}
NES package with examples and tutorials is available at 
\url{https://github.com/sgrubas/NES}.

\bibliographystyle{unsrt}  
\bibliography{references}  

\end{document}

%% file: Introduction.tex
\section{Introduction} \label{Introduction}

The eikonal equation is a first-order non-linear partial differential equation (PDE) that is used in various applications: geometric optics, computer vision, image processing and others \cite{sethian1999level}. It is also used for computing the first-arrival traveltimes of seismic waves in many seismic applications: ray modeling \cite{abgrall1999big}; ray multipathing \cite{rawlinson2010multipathing}; traveltime tomography \cite{schuster1993wavepath}; Kirchhoff migration \cite{gray1994kirchhoff}. There are well-known numerical techniques for solving the equation: Fast Marching Method \cite{sethian1999fast, treister2016fast} and Fast Sweeping Method  \cite{zhao2005fast, fomel2009fast}. Often, the seismic applications require massive computations of the traveltimes for many source-receiver pairs, demanding a high accuracy in complex velocity models. 

The Physics-Informed Neural Networks (PINNs) were introduced for solving PDEs via artificial neural networks. To date, PINNs are used for many equations including Schrodinger, Burgers, Navier-Stocks and some others \cite{sirignano2018dgm, raissi2019physics}.
PINNs are trained by minimizing a loss function which contains the governing PDE, corresponding boundary and initial conditions. The input data to the PINNs consist of collocation points in the domain of PDE. During the training, the neural network learns the PDE by minimizing the loss function on the training points in the domain of interest, giving a grid-free approximate solution to the PDE.  
Many enhancements for the PINNs have been proposed to improve its performance: adaptive activation \cite{jagtap2020adaptive, jagtap2020locally}; gradient of PDE \cite{yu2022gradient}; loss balancing \cite{bischof2021multi, wang2021understanding}; improved architectures \cite{wang2021understanding}; adaptive sampling \cite{lu2021deepxde, nabian2021efficient}. Some of the improvements almost do not increase computational cost, but others may considerably slow down the training. 

PINNs have already been used for solving the eikonal equation \cite{smith2020eikonet, grubas2020seismic, bin2021pinneik}. Existing implementations of PINNs for the eikonal equation have low accuracy in the presence of caustics or require significant computational resources for the training. In this paper, we consider further development of a robust PINN design for solving the eikonal equation.
The usage of PINN-based eikonal solver has several advantages in the seismic applications. For example, PINNs provide not only the traveltimes but the derivatives, which may speed up the solving of inverse problems: localization of earthquakes \cite{smith2021hyposvi} and microseismic events \cite{grubas2021localization}; traveltime tomography \cite{silvaphysics, waheed2021pinntomo}. Other advantage is compression of the traveltime tables for seismic imaging problems \cite{grubas2020traveltime}.

In this paper, we propose the novel Neural Eikonal Solver (NES) for solving the eikonal equation using the PINNs. 
First, we introduce a new version of factorizing the eikonal equation. It accounts for the singularity at the source and limits the classes of functions satisfying the eikonal equation. Second, we propose a new form of the loss function to account for the errors in caustic regions. Third, we propose the activation function for better approximation of the caustic regions. Fourth, we introduced a symmetrization of the solution to account for the reciprocity principle. Lastly, we proposed the input scaling for invariance to the units (m/s or km/s) and the random weight initialization for speeding up the convergence. The proposed techniques do not increase the computational overhead and are accompanied by systematic testing in different velocity models. 
In comparison to other PINN-based eikonal solvers the NES performance shows a higher traveltime accuracy and a faster training for an arbitrary velocity model. 


Our paper is organized as follows: in chapter \ref{Theory} we introduce the eikonal equation and our new PINN architecture (NES); in chapter \ref{Testing} we test the NES performance for different velocity models and provide comparison with existing solutions; in chapter \ref{Discussion} we provide discussion of future developments and potential applications; chapter \ref{Conclusions} summarises results and concludes our paper.

%% file: Theory.tex
\section{Theory} \label{Theory}


\subsection{Eikonal equation and neural networks}
The eikonal equation is used for computing traveltimes of seismic waves in a high-frequency approximation. For isotropic heterogeneous velocity models the equation is often written in the following form \cite{cerveny2001seismic}:
\begin{equation}
    \begin{array}{c}
         \left\Vert \nabla_r  \tau(\textbf{x}_r) \right\Vert=v(\textbf{x}_r)^{-1}, \\
         \tau (\textbf{x}_s) = 0, 
    \end{array}
    \label{eq:OnePointEikonal}
\end{equation}
where $\Vert \cdot \Vert$ is the Euclidean norm, $\nabla_r \equiv (\partial_{x_r}, \partial_{y_r}, \partial_{z_r})$ is the gradient operator w.r.t. to coordinate $\textbf{x}_r=(x_r, y_r, z_r)$, $v(\textbf{x}_r)$ is a velocity model. This equation defines the first-arrival traveltimes for a fixed source location $\textbf{x}_s=(x_s, y_s, z_s)$, constrained by the boundary condition $\tau (\textbf{x}_s) = 0$. There are several numerical techniques for solving the equation in a stable manner, e.g. Fast Marching Method (FMM) \cite{sethian1999fast} and Fast Sweeping Method (FSM) \cite{zhao2005fast}. 

In seismic applications it is commonly needed to compute traveltimes for multiple sources \cite{claerbout1985imaging, goldin1986seismic}. Therefore, we can consider traveltime field as function of two points $T(\textbf{x}_s, \textbf{x}_r)$ that should satisfy the reciprocity principle for monotype waves:  
\begin{equation}
    \begin{array}{c}
         \left\Vert \nabla_r T(\textbf{x}_s, \textbf{x}_r) \right\Vert=v(\textbf{x}_r)^{-1}, \\ \left\Vert \nabla_s T(\textbf{x}_s, \textbf{x}_r) \right\Vert=v(\textbf{x}_s)^{-1}, \\
         T(\textbf{x}_s, \textbf{x}_r) = T(\textbf{x}_r, \textbf{x}_s), \\
         T(\textbf{x}_s, \textbf{x}_s)=0, 
    \end{array}
\label{eq:TwoPointEikonal}
\end{equation}
where  $\nabla_s \equiv (\partial_{x_s}, \partial_{y_s}, \partial_{z_s})$ and $\nabla_r \equiv (\partial_{x_r}, \partial_{y_r}, \partial_{z_r})$ denote gradients with respect to the source $\textbf{x}_s=(x_s, y_s, z_s)$ and the receiver $\textbf{x}_r=(x_r, y_r, z_r)$ coordinates. With that, $\nabla_s$ and $\nabla_r$ define the source-part and the receiver-part eikonal equations. Unlike the eikonal for a fixed source (one-point form) (eq. \ref{eq:OnePointEikonal}), this system describes the first-arrival traveltimes for all the source-receiver pairs, and we will call it the two-point eikonal.

Physics-Informed Neural Networks (PINN) is a concept for solving PDEs usually by means of using the Fully-Connected Neural Networks (FCNNs) \cite{raissi2019physics}. 
The FCNN is a universal approximator \cite{hornik1989multilayer} representing the mapping $\boldsymbol{f}_{\boldsymbol{\theta}}(\mathbf{x})$: $\mathbb{R}^n \rightarrow \mathbb{R}^m$. FCNN with $K-1$ hidden layers can be written in the following form:
\begin{equation}
    \boldsymbol{h}_k = g_k(\boldsymbol{\Theta}_k \boldsymbol{h}_{k-1} + b_k),\,\,\,\,k=0...K,
    \label{eq:FCNN}
\end{equation} 
where $k$ is a hidden layer index, $h_k \in \mathbb{R}^{N_k \times 1}$ is a hidden state, $N_k$ is a number of units on a hidden layer $k$,  $\boldsymbol{\Theta}_k \in \mathbb{R}^{N_k \times N_{k-1}}$ are kernel weights, $b_k$ is a bias weight, $g_k$ is an activation function providing the non-linearity. For an arbitrary regression problems the last activation $g_K$ is usually the linear function. Note, $\mathbf{h}_0=\mathbf{x} \in \mathbb{R}^{n}$ is the input data, and $\boldsymbol{f}_{\boldsymbol{\theta}}(\mathbf{x}) = \boldsymbol{h}_K \in \mathbb{R}^{m}$ is the output, where $\boldsymbol{\theta}$ denotes all the weights of the network $(\boldsymbol{\Theta}_1, b_1, ..., \boldsymbol{\Theta}_K, b_K)$. The weights $\boldsymbol{\theta}$ are selected during the training procedure, which implies  
adjusting the weights $\boldsymbol{\theta}$ by minimizing the loss function $L(\boldsymbol{\theta})$ with respect to the training data set. This loss function represents the distance between the neural-network output and the training set. Since a neural network $f_{\boldsymbol{\theta}}(\mathbf{x})$ is a nonlinear operator, training is often performed using the optimization methods based on a gradient descent. 

In our work we used the Adam optimization method, which is the most popular and provides relatively fast and accurate convergence for the majority of problems \cite{kingma2014adam, reddi2019amsgrad}. For the problems where the training-set size $M$ is huge, the data are split into small batches to fit the machine memory (mini-batch gradient descent). 
In the computational frameworks such as Tensorflow \cite{abadi2016tensorflow} the algorithm of Automatic Differentiation (AD) is used, which allows computing the analytical derivatives of any order at a machine precision \cite{baydin2018automatic}. This Automatic Differentiation is crucial for PINNs because it allows both constructing a loss function $L(\boldsymbol{\theta})$ for given PDE and computing gradient for its optimization. 


\subsection{New Neural Eikonal Solver (NES)}
We introduce new Neural Eikonal Solver (NES) based on the PINN concept. This approach is mesh-free, unlike the finite-difference methods (FMM, FSM). The target grid of traveltimes can be initialized with an arbitrary distribution of collocation points and will not require any modifications of the algorithm. 
Existing neural-network eikonal solvers \cite{bin2021pinneik,smith2020eikonet} exhibit different performance depending on the complexity of the velocity model. They may require longer training time or accuracy deterioration for complicated velocity models. Main cause for this problems is that the eikonal equation is a non-linear PDE that can produce non smooth front shapes in heterogeneous velocity models producing caustics. 
Thus, we want to revisit the way of constructing the neural-network solver in order to make it fast and reach minimum number of trainable parameters providing high accuracy of the solution for complicated velocity models.  

The NES has two versions: 
\begin{itemize}
  \item solving one-point problem with fixed source (NES-OP), equation \ref{eq:OnePointEikonal};
  \item solving two-point problem for arbitrary source-receiver pair (NES-TP), equation \ref{eq:TwoPointEikonal}.
\end{itemize}


\subsubsection{Improved factorization of the eikonal equation}


The solution to the eikonal equation has singularity at the source point which deteriorates the solution accuracy in this region. To avoid this problem, it was suggested to use factorization \cite{fomel2009fast}, i.e. look for the eikonal solutions in the form (for the one-point and the two-point problem respectively): 
\begin{equation}
    \tau(\textbf{x}_r) = R \cdot f(\textbf{x}_r),
    \label{eq:FactorizationOP}
\end{equation}
\begin{equation}
    T(\textbf{x}_s, \textbf{x}_r) = R \cdot F(\textbf{x}_s, \textbf{x}_r),
    \label{eq:FactorizationTP}
\end{equation}
where $R=\Vert \textbf{x}_r - \textbf{x}_s \Vert$ is the distance function and the solution in an isotropic homogeneous velocity model ($v=1$), $f(\textbf{x}_r)$ and $F(\textbf{x}_s, \textbf{x}_r)$ are unknown functions representing deviations from the solution in homogeneous model (for the one-point and the two-point eikonal respectively). The factor $R$ emulates the source-point singularity, which significantly enhances the accuracy of FMM \cite{treister2016fast} and FSM \cite{fomel2009fast}. This factorization is also used in the neural-network eikonal solvers \cite{smith2020eikonet,bin2021pinneik}. 

\begin{figure}[!htb]
\centering
    \includegraphics[width=0.5\textwidth]{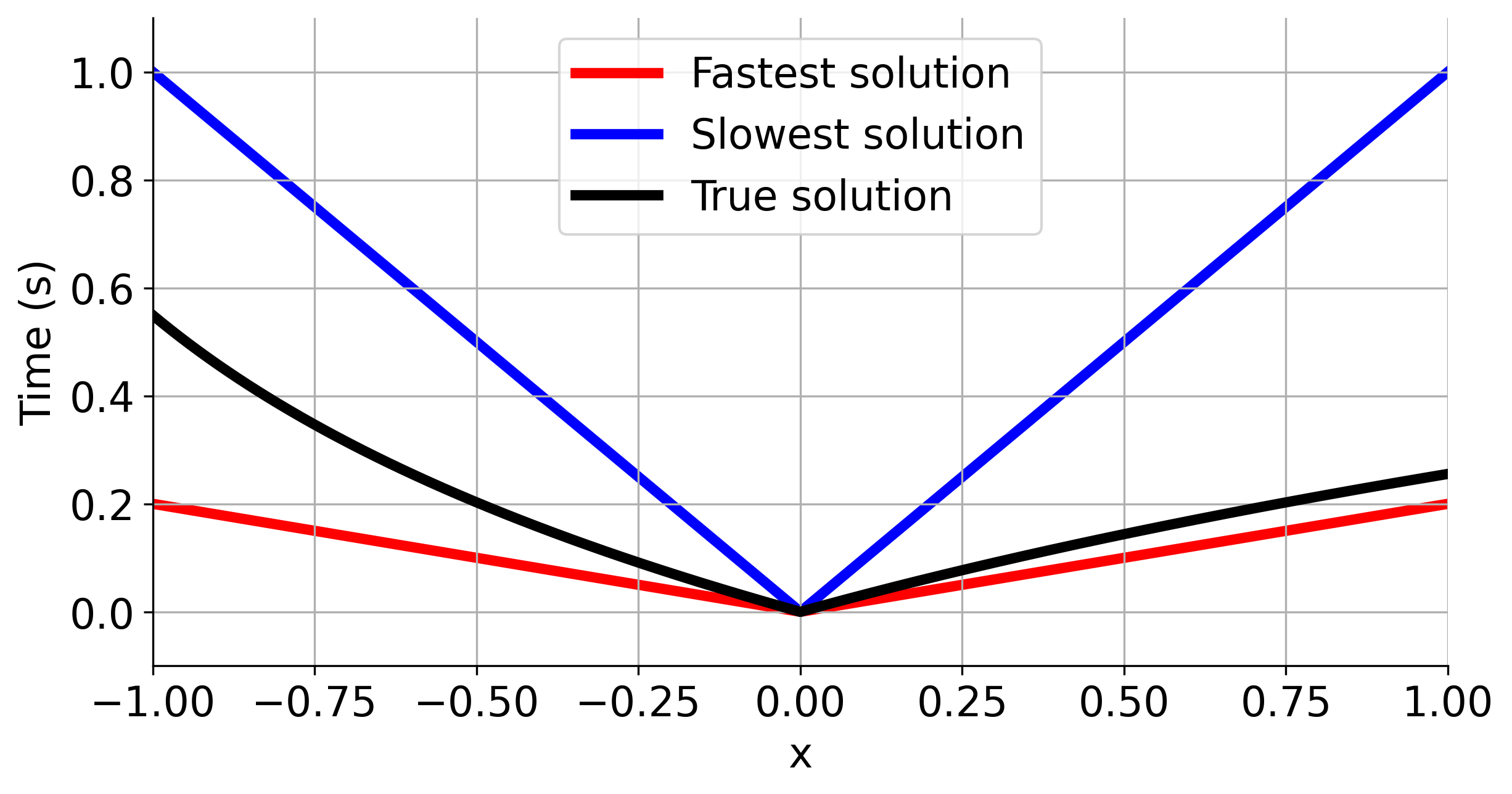}
   \caption{Solution of the eikonal equation for 1D velocity model with a linear gradient. Black line indicates the true solution which is bounded by the slowest (blue line) and the fastest (red line) solutions.}
   \label{fig:Factorization}
\end{figure}

If we use PINNs for the eikonal solution, the functions $f(\textbf{x}_r)$ and $F(\textbf{x}_s, \textbf{x}_r)$ are to be approximated. However, their free form requires PINN to search the solution from the huge class of functions which may be time consuming resulting in the long training. Therefore, we propose the improved factorization which limits the class of functions for search, thereby significantly boosting the training speed of the NES.
Let us consider the following expression for the NES-OP and NES-TP solutions respectively:
\begin{equation}
     \tau_{\boldsymbol{\theta}}(\textbf{x}_r) = R \cdot \bar{\sigma}\left(f_{\boldsymbol{\theta}}(\textbf{x}_r)\right),
    \label{eq:ImpFactorizationOP}
\end{equation}
\begin{equation}
     T_{\boldsymbol{\theta}}(\textbf{x}_s, \textbf{x}_r) = R \cdot \bar{\sigma}\left(F_{\boldsymbol{\theta}}(\textbf{x}_s, \textbf{x}_r)\right),
    \label{eq:ImpFactorizationTP}
\end{equation}
where $f_{\boldsymbol{\theta}}$ and $F_{\boldsymbol{\theta}}$ are the neural-network outputs, $\bar{\sigma}$ is a bounding function or the output activation function ($g_{K}$ in eq. \ref{eq:FCNN}). The NES solution must be strictly positive function $\tau_{\boldsymbol{\theta}}(\textbf{x}_r) > 0$ $\forall\, \textbf{x}_r \neq \textbf{x}_s$ (same for $T_{\boldsymbol{\theta}}$). Since $R \geqslant 0$, then the second term must always be strictly positive $\bar{\sigma}>0$. Note, $R$ is the homogeneous solution for $v=1$ and $\bar{\sigma}$ is supposed to be a small deviation from the homogeneity. If we assume that the unity velocity is the minimal possible value, so then $R$ constrains the range of $\bar{\sigma}$ to be in $(0, 1]$. Any velocity model $v$ can be represented using its ranges: $v_{min} < v < v_{max}$, where $v_{min}, v_{max}$ are minimal and maximal values in a considered velocity model respectively. This bounding allows to constrain $\bar{\sigma}$ in the range of $[1/v_{max}, 1/v_{min}]$. It bounds the NES solution to be $\tau_{\boldsymbol{\theta}},T_{\boldsymbol{\theta}} \in [R/v_{max}, R/v_{min}]$, i.e. constrains the NES between the fastest and the slowest solutions. We suggest the following bounding function $\bar{\sigma}$:
\begin{equation}
     \bar{\sigma}(x) = \displaystyle \left(\frac{1}{v_{min}} - \frac{1}{v_{max}}\right) \sigma(x) + \frac{1}{v_{max}},
     \label{eq:secondTerm}
\end{equation}
where $\sigma$ may be any function defined in the range of $[0,1]$. We propose the sigmoidal function because it showed the best performance:
\begin{equation}
     \sigma(x) = \frac{1}{1 + e^{-x}},
     \label{eq:sigmoid}
\end{equation} 
The explanation of the improved factorization is shown in Figure~\ref{fig:Factorization}, where the true solution is always bounded by the slowest and fastest solutions. That also means that if a velocity model is homogeneous $v_{min}=v_{max}=v$ then the NES converges to the true solution, i.e. $\tau_{\boldsymbol{\theta}}=R/v$ everywhere (same for $T_{\boldsymbol{\theta}}$). 


\subsubsection{Imposing reciprocity principle}
The solution to the two-point eikonal \ref{eq:TwoPointEikonal} can benefit when based on the reciprocity principle. The reciprocity $T_{\boldsymbol{\theta}}(\textbf{x}_s, \textbf{x}_r)=T_{\boldsymbol{\theta}}(\textbf{x}_r, \textbf{x}_s)$ can be introduced as an additional constraint to the loss function or within the factorization. The first approach is a soft constraint which does not guarantee the reciprocity. The soft constraint means that it is introduced through the penalty so that the constraint is not strictly satisfied. The second approach is a hard constraint which can guarantee reciprocity. The hard constraint means that it is strictly satisfied through imposing a certain behaviour on the function. We suggest to use the hard type of constraint to guarantee the desirable solution behaviour. The reciprocity principle requires constructing a symmetric function with respect to permutation of the two arguments $\textbf{x}_s$ and $\textbf{x}_r$, which can be done by including the averaging of the permutations in the NES-TP approximation:
\begin{equation}
     T_{\boldsymbol{\theta}}(\textbf{x}_s, \textbf{x}_r) = R \cdot \bar{\sigma}\left(\frac{F_{\boldsymbol{\theta}}(\textbf{x}_s, \textbf{x}_r) + F_{\boldsymbol{\theta}}(\textbf{x}_r, \textbf{x}_s)}{2}\right).
    \label{eq:ImpFactorizationReciprocityTP}
\end{equation}
This guarantees reciprocity everywhere $T_{\boldsymbol{\theta}}(\textbf{x}_s, \textbf{x}_r) \equiv T_{\boldsymbol{\theta}}(\textbf{x}_r, \textbf{x}_s)$ and combines the receiver- and source-part equations (eq. \ref{eq:TwoPointEikonal}) into a single one. The last assumption is valid as the gradients $\nabla_r T_{\boldsymbol{\theta}}$ and $\nabla_s T_{\boldsymbol{\theta}}$ for \ref{eq:ImpFactorizationReciprocityTP} define the same function but measured at different spatial coordinates $(\textbf{x}_s, \textbf{x}_r)$ and $(\textbf{x}_r, \textbf{x}_s)$ respectively. Thus, the expression \ref{eq:ImpFactorizationReciprocityTP} provides the required reciprocity and only one equation is needed to solve the two-point eikonal system \ref{eq:TwoPointEikonal}. For the simplicity we will consider only the receiver-part equation for the NES-TP hereafter. 

\subsubsection{Form of the loss function}

Here we discuss a proper choice of the loss-function to achieve better training results for the NES. 
The factor $R$ in
\ref{eq:ImpFactorizationOP} and \ref{eq:ImpFactorizationTP}
automatically satisfies the boundary condition allowing to exclude it from the loss function. Thus, we need to use only the eikonal equation while forming the loss function in NES. 
Note that the eikonal equation is zero-level set of a corresponding Hamiltonian that may take various forms \cite{cerveny2001seismic}. In particular, we can write for the one-point \ref{eq:OnePointEikonal} and the two-point formulations \ref{eq:TwoPointEikonal}
correspondingly: 
\begin{equation}
    \mathcal{H}_p(\textbf{x}_r, \tau) = \textstyle \frac{1}{p}\left[v(\textbf{x}_r)^p \Vert\nabla \tau(\textbf{x}_r)\Vert^p - 1\right] = 0,
    \label{eq:HamiltonianOP}
\end{equation}
\begin{equation}
\begin{array}{c}
     \mathcal{H}_p(\textbf{x}_s, \textbf{x}_r, T) =\frac{1}{p}\left[v(\textbf{x}_r)^p \Vert\nabla_r T(\textbf{x}_s, \textbf{x}_r)\Vert^p - 1\right] = 0, 
\end{array}
\label{eq:HamiltonianTP}
\end{equation}
where $p \neq 0$. 

So we have a family of possible loss functions depending on the parameter $p$ (for the one-point and the two-point formulations respectively):
\begin{equation}
     L_{op}(\boldsymbol{\theta}, p) = \frac{1}{N_r} \displaystyle \sum_{\textbf{x}_r} \left| \mathcal{H}_p(\textbf{x}_r, \tau_{\boldsymbol{\theta}}) \right|,
    \label{eq:LossOP}
\end{equation}
\begin{equation}
     L_{tp}(\boldsymbol{\theta}, p) = \frac{1}{N_{sr}} \displaystyle \sum_{\textbf{x}_s, \textbf{x}_r} \left| \mathcal{H}_p(\textbf{x}_s, \textbf{x}_r, T_{\boldsymbol{\theta}}) \right|,
    \label{eq:LossTP}
\end{equation}
where subscripts $_{op}$ and $_{tp}$ refer to the one-point and two-point eikonals respectively, $N_r$ is a number of receiver points, $N_{sr}$ is a number of source-receiver pairs, $|\cdot|$ denotes an absolute value, $\mathcal{H}_p$ refers to a Hamiltonian form (eqs. \ref{eq:HamiltonianOP} and \ref{eq:HamiltonianTP}). These loss functions are based on the $L_1$-norm (mean-absolute error) of the Hamiltonian. Below we explain the advantages of the proposed loss function.

\textit{Loss function for robust solution}

In many PINNs application for different PDEs the $L_2$-norm (mean-squared error) in the loss function has been used \cite{raissi2019physics, jagtap2020adaptive, jagtap2020locally, yu2022gradient,
bischof2021multi, wang2021understanding, lu2021deepxde, nabian2021efficient} including the eikonal equation \cite{smith2020eikonet, bin2021pinneik}. The $L_2$-norm is designed for a normal Gaussian distribution of the errors. In case of the eikonal equation the error distribution may not be Gaussian for an arbitrary heterogeneous velocity models, especially when the eikonal solution may have caustics. At the caustics the seismic-rays trajectories intersect each other forming singularity zone where the gradients are discontinuous and undefined, that is, the equation is not satisfied. In vicinity of the caustic singularities the solution itself is not stable which leads to the presence of errors. Unlike the source-point singularity, the caustics singularities cannot be localized in advance. Therefore, we have to account for the errors related to the caustic singularities using other type of error metric. In regression problems, the $L_1$-norm (mean-absolute error) is known to provide the solution which is robust to outliers, assuming that the error has random Laplace distribution \cite{rousseeuw2005robust}. That is why, we suggest to use $L_1$-norm (mean-absolute error) in the loss functions \ref{eq:LossOP} and \ref{eq:LossTP}. 

\textit{Non-symmetric error distribution}

\begin{figure*}[!htb]
    \centering
    \begin{subfigure}[b]{0.4\textwidth}
         \centering
         \includegraphics[width=\textwidth]{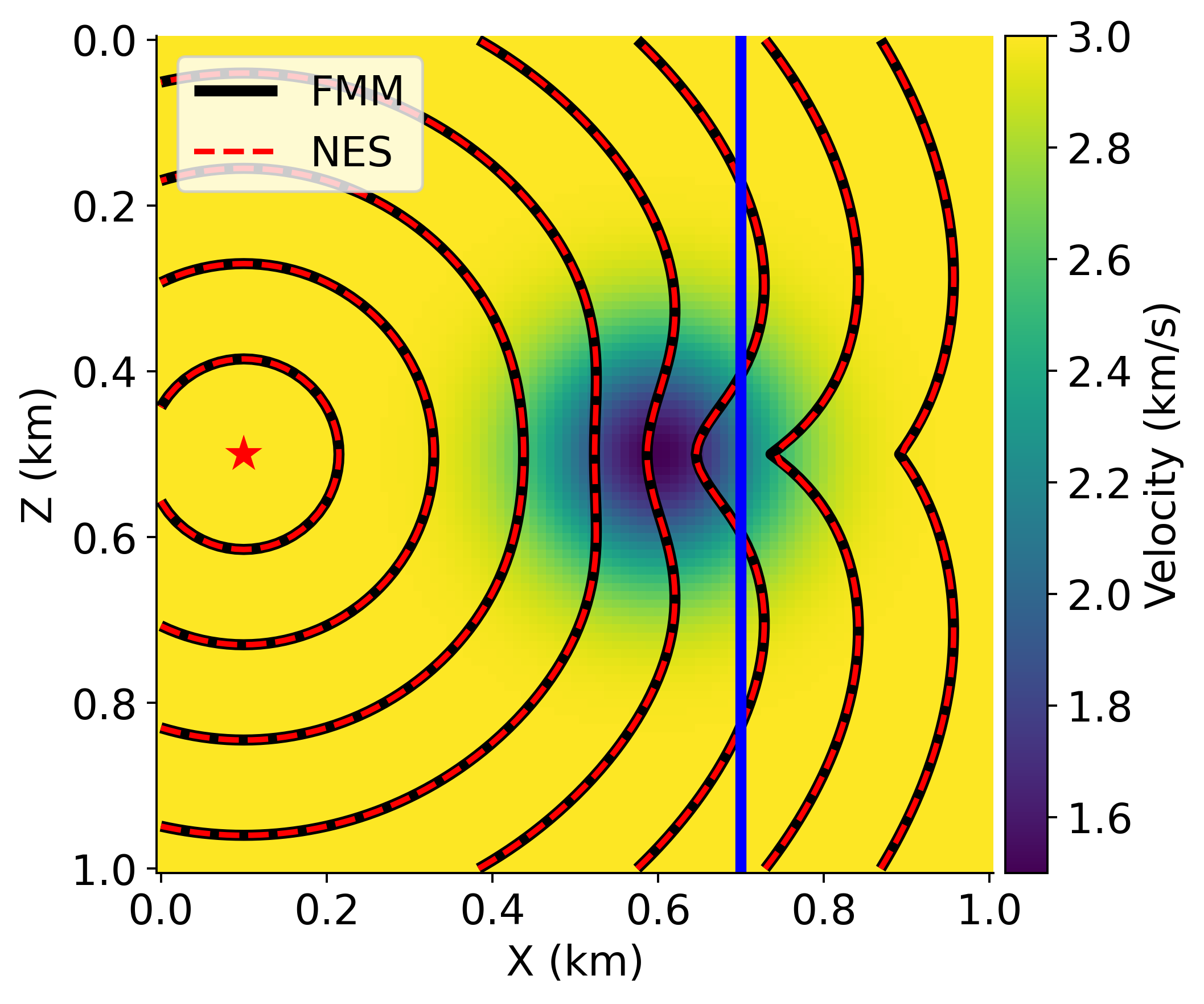}
         \caption{}
         \label{fig:LossFunction_Nonsymmetry_Explain_Model}
     \end{subfigure}
    \hfill
    \begin{subfigure}[b]{0.5\textwidth}
         \centering
         \includegraphics[width=\textwidth]{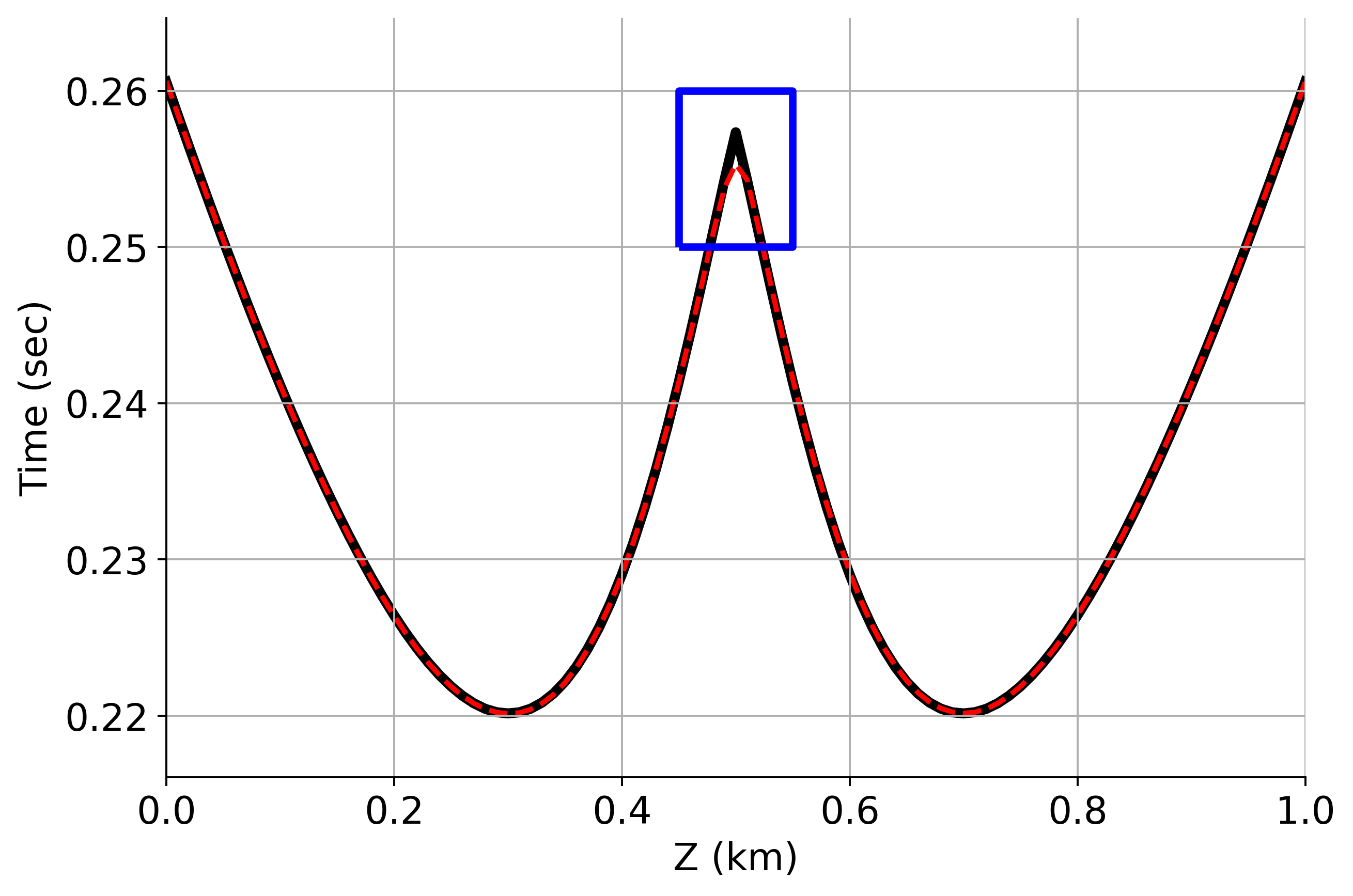}
         \caption{}
         \label{fig:LossFunction_Nonsymmetry_Explain_Tsection}
     \end{subfigure}
     \caption{The example of the caustic singularities. \textbf{(a)} shows the velocity model, where red star indicates the source position, the black solid contours represent the reference solution (the second-order factored FMM), the red dashed contours represent the NES-OP solution, the blue vertical line at $x=0.7$ km indicates a section for \textbf{(b)}. \textbf{(b)} showcases the vertical section of the solutions where the blue box highlights the position of the caustic singularity. From \textbf{(b)} we can see that the NES (red dashed line) smooths the caustic singularity (in blue box) at $z=0.5$ km.}
    \label{fig:LossFunction_Nonsymmetry_Explain}
\end{figure*}

The errors due to the caustic singularities have a non-symmetric distribution. As we mentioned, the eikonal solution tends to be unstable at the caustic singularities which are the zones of a non-smooth behaviour of a solution. The NES approximation, which is a smooth function, tends to smooth this zone (see Figure~\ref{fig:LossFunction_Nonsymmetry_Explain}). From the gradient of the solution we can obtain a predicted velocity model in a form of $v_{\boldsymbol{\theta}} = \Vert\tau_{\boldsymbol{\theta}}\Vert^{-1}$, and the difference with the actual velocity model $v$ may show how well the eikonal equation is satisfied. The smoothing at caustic singularities means that the NES approximation has a lower magnitude of the solution gradients (Figure~\ref{fig:LossFunction_Nonsymmetry_Explain_Tsection}), that is, the predicted velocity exceeds the actual velocity values ($v_{\boldsymbol{\theta}}>v$). Thus, the errors related to the caustic singularities are shifted to the zone of exceeding velocity values. This observation was confirmed with the internal experiments. Therefore, we should allow the high-velocity errors to exist with the higher probability than the low-velocity errors which do not represent caustics. To do so, we should make our Laplace distribution non-symmetric so that the high-velocity errors have higher probability. 

The Hamiltonian form (eqs. \ref{eq:HamiltonianOP} and \ref{eq:HamiltonianTP}) can help to take a non-symmetric distribution into account. The Hamiltonian $\mathcal{H}_p$ uses the ratio of the predicted and actual velocities (the term $v^p \Vert\tau_{\boldsymbol{\theta}}\Vert^p$ which is equivalent to $v^p /  v_{\boldsymbol{\theta}}^p$). If the predicted velocity is higher than the actual one, the ratio is less than unity $v^p /  v_{\boldsymbol{\theta}}^p<1$. It means that we can reduce the ratio magnitude when it is less than unity (decrease the significance of high-velocity errors) and enhance when it is higher (increase the significance of low-velocity errors) by setting up the $p$ value to be > 1. Our tests showed that the optimal value is $p=2$ (in extreme cases with many caustics it can be $p=3$). Noteworthy, the Hamiltonian $\mathcal{H}_p$ with $p=2$ defines the ray-tracing system parameterized by traveltime \cite{cerveny2001seismic}.

\textit{Non-dimensional loss}

The Hamiltonian-based loss function (eqs. \ref{eq:LossOP} and \ref{eq:LossTP}) is non-dimensional as it is scaled by the actual velocity model. The loss function becomes non-sensitive to the velocity model and may give the same loss values for the velocity models of different order of values. It allows setting a single tolerance value as a training conditioning for the loss function for any velocity model.

\subsubsection{Activation function} \label{Activation}

The choice of the hidden activation $g_k$ (eq. \ref{eq:FCNN}) is important because it describes the basis functions for solution approximation. All the functions used for PINN should be smooth enough to provide the existence of the necessary derivatives. The neural-network output is differentiated w.r.t. inputs (spatial coordinates) in computing the losses (eqs. \ref{eq:LossOP} and \ref{eq:LossTP}) and also w.r.t. weights during the training procedure. 
Thus, we need to consider at least two-times differentiable activation functions.

The shape of the caustic singularities in Figure~\ref{fig:LossFunction_Nonsymmetry_Explain_Tsection} (at the center $z=0.5$ km) are smoothed by the NES approximation. We suggest to approximate these singularities by a set of the gaussian functions $g_k(x)=e^{-x^2}$ because it has natural extremum resembling the singularity shape and also is infinitely differentiable function. The gaussian function is also called as Radial Basis Function which are universal approximators \cite{park1991universal}. 
In all our tests, the $gauss$ activation showed the significantly better performance than the other activation functions for the velocity models with caustics. 

\subsubsection{Other neural-network parameters} \label{Hypers}

There are many hyperparameters of the neural networks that may influence the performance of a training and accuracy of a result. The common practice is to test the different set of parameters and select it based on the best performance \cite{omalley2019kerastuner}. In this subsection we consider the choice of initial weights, input data scaling, and the type of connection between hidden layers. The discussed parameters are chosen among the other results of our research based on the most significant impact on the accuracy and performance.

The weights of neural network must be initialized before the training. These initial weights are used to be initialized in random manner, but the type of the distribution can be different. We tested different types of distribution ($\boldsymbol{\Theta}_k$ in equation \ref{eq:FCNN}) which are available in Keras framework \cite{chollet2015keras}. The $he\_normal$ and $he\_uniform$ \cite{he2015delving} types of random distributions provided a considerably faster convergence of the loss function, which starts decreasing from the first epochs and almost without any stagnation. 

The input data scaling or normalization is another popular technique to improve the training process of neural networks: considerably speeds up the convergence of the gradient descent and standardizes the learning rate definition. We propose the scaling of the inputs $\textbf{x}_r$ (and $\textbf{x}_s$) by $\max|\textbf{x}_r|$ (or $\max(|\textbf{x}_s|, |\textbf{x}_r|)$) which maps input data to the range of $[-1, 1]$. Such a scaling prevents the neural network to be sensitive to the units of the spatial coordinates ($m$ or $km$). 

We have considered several types of architectures of neural networks and types of connection between the hidden layers. The fully-connected layers provide the best performance compared to the layers with the skip-connections (residual blocks) \cite{huang2018resnet} and attention mechanism \cite{wang2021understanding}. We tested the influence of an architecture complexity on the solution accuracy (number of hidden layers and units). Our tests showed that the total number of weights $10^3-10^4$ is enough to provide the accuracy of $0.1\%$, and more weights $> 10^4$ do not give any improvement. From this range, we chose 4 hidden layers with 75 units on each as a baseline because it provides the faster convergence with the least number of weights. However, one still can use less weights for simple velocity models, and more weights for complex velocity models. 

\subsubsection{Final architectures}
\begin{figure*}[!htb]
    \centering
    \begin{subfigure}[b]{0.825\textwidth}
          \centering
          \includegraphics[width=\textwidth]{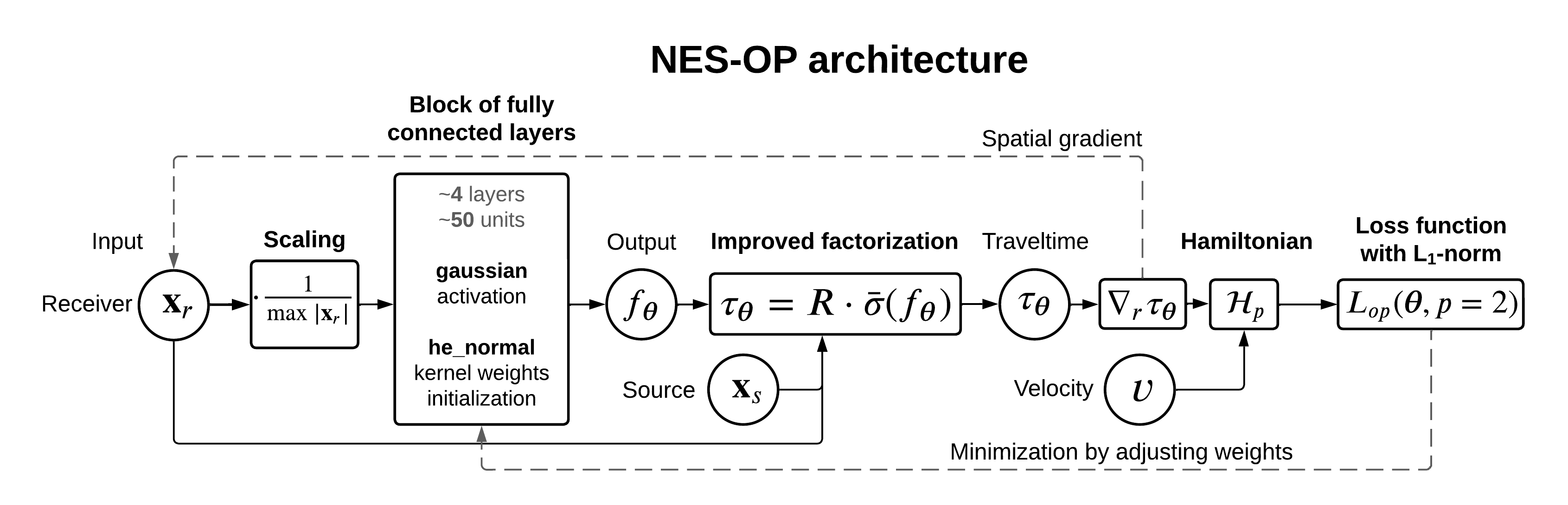}
          \caption{}
          \label{fig:NESOP_Architecture}
    \end{subfigure}
    \hfill
    \begin{subfigure}[b]{\textwidth}
          \centering
          \includegraphics[width=\textwidth]{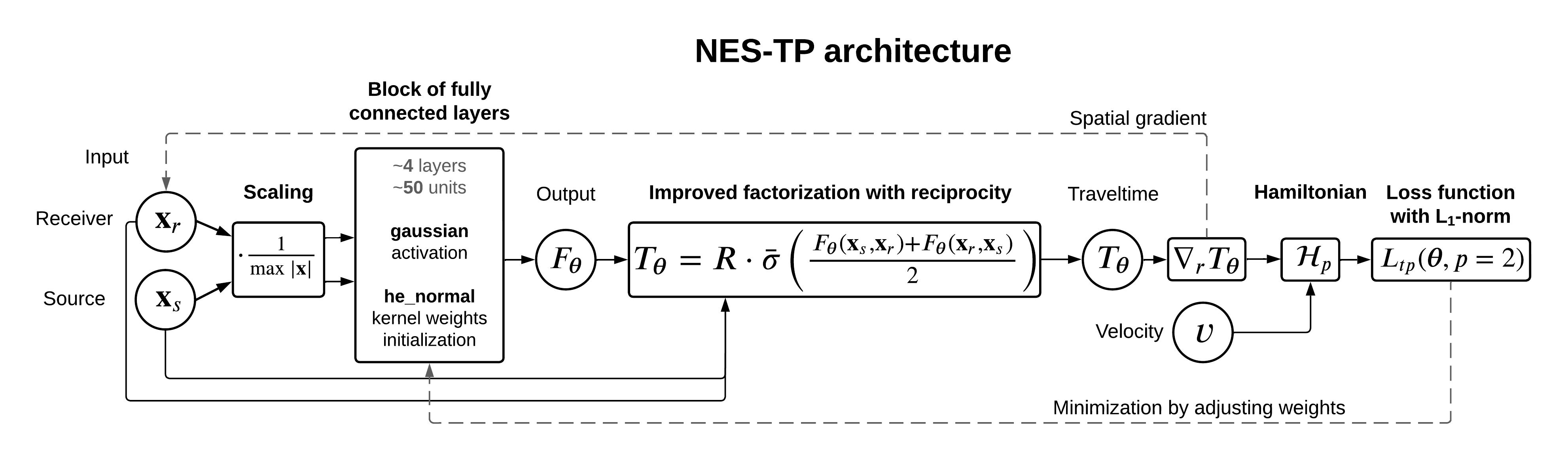}
          \caption{}
          \label{fig:NESTP_Architecture}
     \end{subfigure}
     \caption{The diagrams of the NES-OP \textbf{(a)} and the NES-TP \textbf{(b)} architectures. The text in bold indicates our main contribution. The gray dashed arrow in the spatial gradient means that traveltime is differentiated with respect to the $\textbf{x}_r$. The gray dashed arrow from the loss function indicates the backpropagation algorithm for the training. The symbol $\textbf{x}$ in the scaling layer in \textbf{(b)} indicates that the maximum value is calculated for both source and receiver locations.}
    \label{fig:NES_Architecture}
\end{figure*}

We propose the architectures for the NES-OP and NES-TP which are visualized in Figure~\ref{fig:NES_Architecture}. The NES-OP approximates the equation for a fixed source location, when a finer grid of collocation points can be used to train for a smaller time to get a very accurate solution. The NES-TP provides the generalized solution for any source-receiver pair. It means that we may need more collocation points and training time, but if we need the solution for many sources, it is recommended to use the NES-TP. Our general recommendation is to use the NES-TP unless a special accuracy is needed. It is also worth noting that the NES package is written using Tensorflow backend \cite{abadi2016tensorflow} and Keras API \cite{chollet2015keras}, which supports parallelism on any heterogeneous systems with single or multiple CPUs, GPUs, or TPUs.

%% file: Testing.tex
\section{Numerical testing} \label{Testing}

In this chapter, we provide: the testing of the proposed NES architectures (Figure~\ref{fig:NES_Architecture}) in the velocity model from Figure~\ref{fig:LossFunction_Nonsymmetry_Explain_Model}; the benchmark and comparison of the NES-OP and the NES-TP performance in Marmousi model; the comparison of the NES with the existing packages PINNeik and EikoNet \cite{bin2021pinneik, smith2020eikonet}. All error estimations are based on the reference solution represented by the second-order factored FMM \cite{treister2016fast} implemented in the package $eikonalfm$ \cite{eikonalfm}. The used error metric is relative mean absolute error (RMAE):
\begin{equation}
RMAE = \frac{\sum |\tau_{ref} - \tau_{\boldsymbol{\theta}}|}{\sum |\tau_{ref}|},
\label{eq:RMAE}    
\end{equation}
where $\tau_{ref}$ is the reference solution, $\tau_{\boldsymbol{\theta}}$ is the NES-OP approximation and can be replaced with $T_{\boldsymbol{\theta}}$ for the NES-TP. The summation is performed for the set of points where we test the accuracy.
\subsection{Advantage of NES architecture}
In this section we show the influence of the proposed NES architectures (Figure~\ref{fig:NES_Architecture}). The velocity model for the test is presented in Figure~\ref{fig:LossFunction_Nonsymmetry_Explain_Model}. The NES-OP was trained on the grid of $61\times61-1$ evenly spaced receivers without the source location because of the singularity. The source is located at the same position as shown in Figure~\ref{fig:LossFunction_Nonsymmetry_Explain_Model}. The RMAE for the NES-OP was measured on $101\times101$ points (2.7 times finer grid). The NES-TP was trained on 25000 collocation points of source-receiver pairs randomly and uniformly distributed in the domain of the same velocity model. The RMAE for the NES-TP was measured on a set of $3\times3$ sources evenly spaced in the domain with $101\times101$ receivers for each. Batch size for both cases was $1/4$ of the training set and Adam method for optimization was used \cite{kingma2014adam} with the learning rate of 2.5$\cdot10^{-3}$. For both scenarios, the neural networks contain 4 hidden layer with 75 units on each layer. 

\begin{figure*}[!htb]
    \centering
    \begin{subfigure}[b]{0.49\textwidth}
          \centering
          \includegraphics[width=\textwidth]{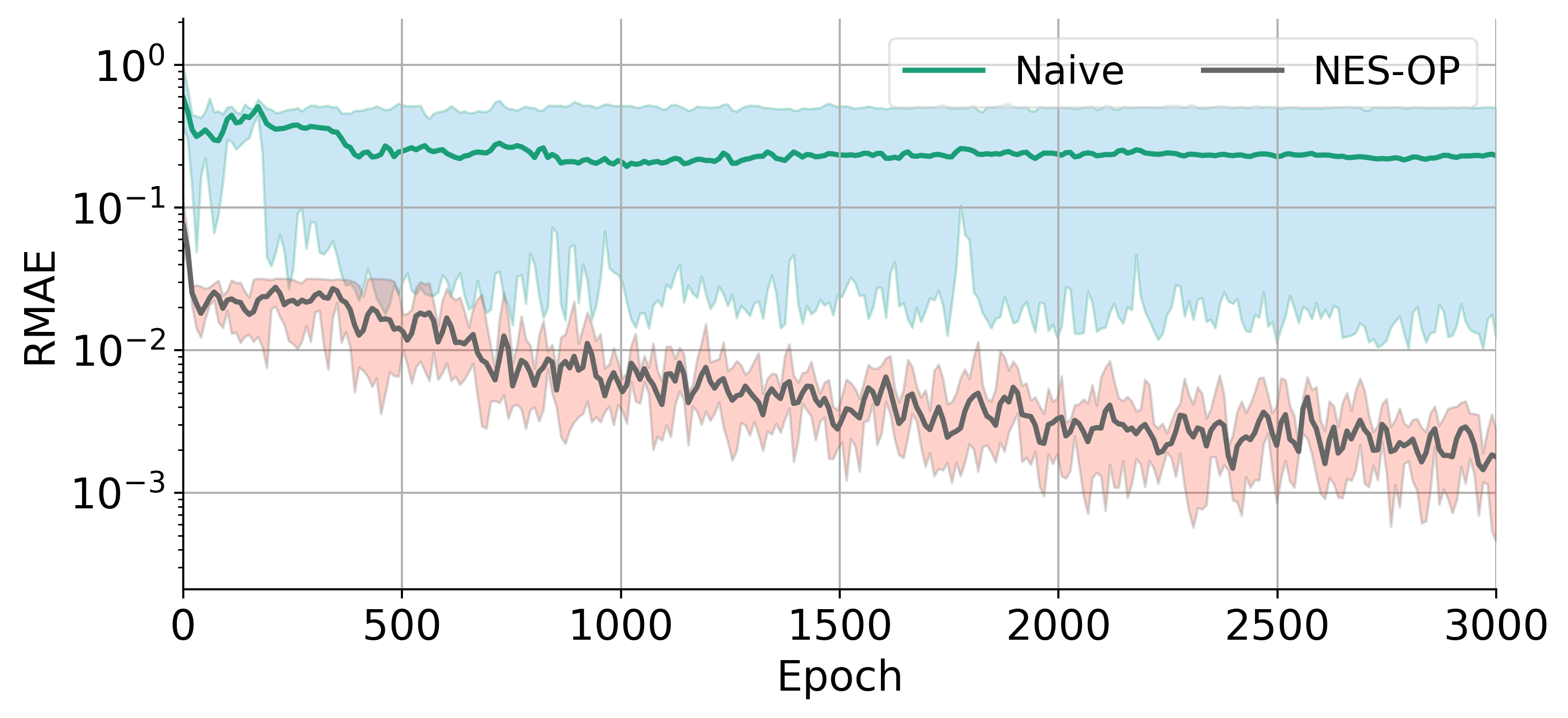}
          \caption{}
          \label{fig:NESOP_LocLowSummary}
    \end{subfigure}
    \hfill
    \begin{subfigure}[b]{0.49\textwidth}
          \centering
          \includegraphics[width=\textwidth]{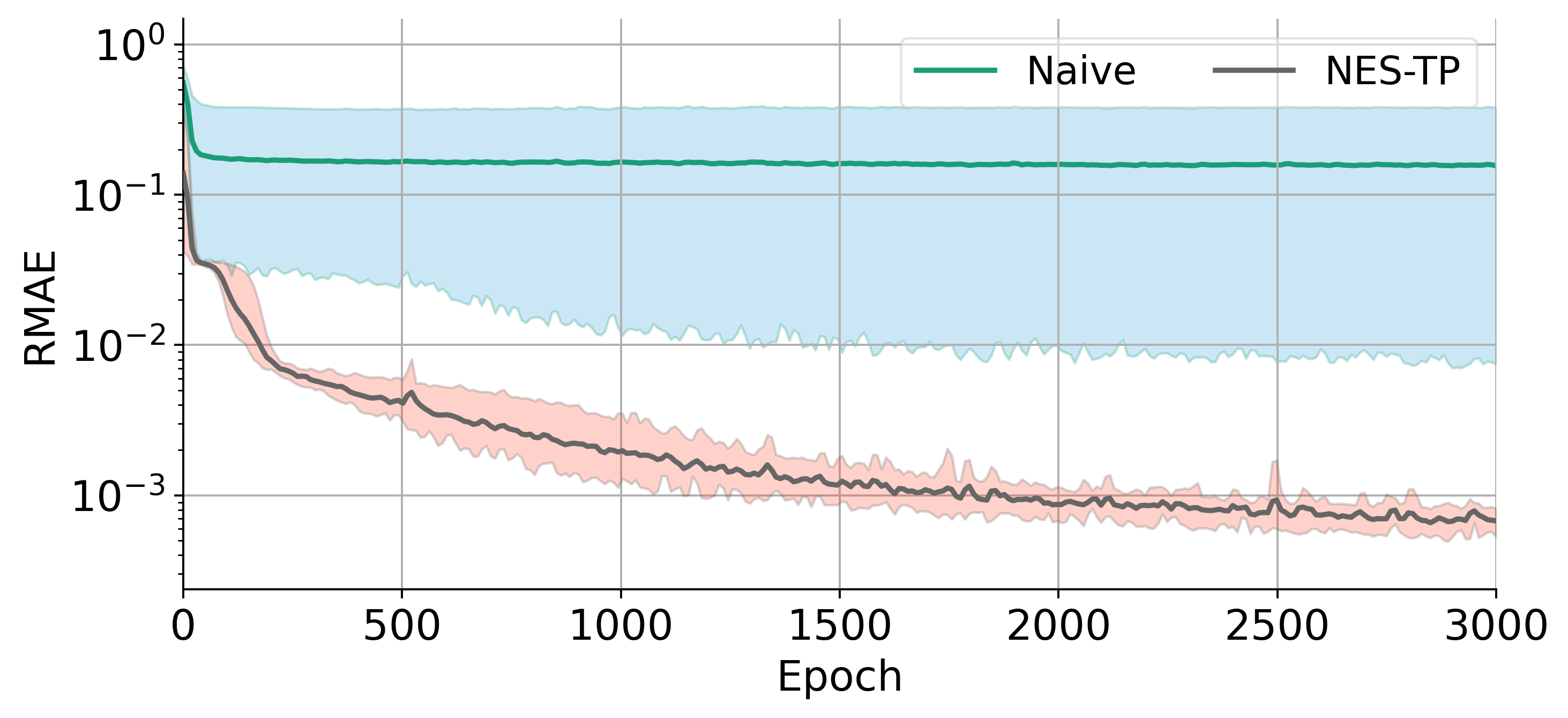}
          \caption{}
          \label{fig:NESTP_LocLowSummary}
     \end{subfigure}
     \caption{The cumulative influence of the NES architectures on the training dynamic in the velocity model from Figure~\ref{fig:LossFunction_Nonsymmetry_Explain_Model}. \textbf{(a)} shows results for the NES-OP architecture (Figure~\ref{fig:NESOP_Architecture}). 
     \textbf{(b)} indicates results for the NES-TP architecture (Figure~\ref{fig:NESTP_Architecture}). Lines indicate average value of RMAE over 5 independent launches of NES, transparent areas around lines indicate min-max band.}
    \label{fig:LocLowSummary}
\end{figure*}

We test the cumulative influence of the proposed NES architectures (all highlighted features by bold font in Figure~\ref{fig:NES_Architecture}). It includes the improved factorization (eqs. \ref{eq:FactorizationOP} and \ref{eq:FactorizationTP}), the loss function with a $L_1$-norm of the Hamiltonian form with $p=2$ (eqs. \ref{eq:LossOP} and \ref{eq:LossTP}), gaussian activation of the hidden layers (subsection \ref{Activation}), the $he\_normal$ weights initialization, the input data scaling (subsection \ref{Hypers}), and the reciprocity principle (eq. \ref{eq:ImpFactorizationReciprocityTP}). We show only cumulative effect because the influence of each component may differ for various set of other settings in the architecture. For example, influence of the gaussian activation with using improved factorization differs from one without the improved factorization. The systematic tests for various models show that the contributions of all the proposed features are approximately equal. The results are shown in Figure~\ref{fig:LocLowSummary}. It shows the training dynamics of the RMAE. Naive refers to an architecture without the improved factorization, with the loss function based on $L_2$-norm and with conventional forms of eikonal equations (eqs. \ref{eq:OnePointEikonal} and \ref{eq:TwoPointEikonal}), with hyperbolic tangent as hidden activation, and without the symmetrization for the reciprocity of the NES-TP. It is clearly seen that the RMAE with the NES architectures is drastically decreased by more than two orders of magnitude and the stability improved (narrower min-max bands). Overall, after 3000 epochs the RMAE was about 0.16\% and 0.07\% for the NES-OP and the NES-TP respectively. These tests were performed on the GPU RTX 3070 Laptop which took 43 and 75 seconds for the training of the NES-OP and the NES-TP respectively.

\subsection{Benchmark on Marmousi 2D}

\begin{figure*}
    \centering
    \includegraphics[width=\textwidth]{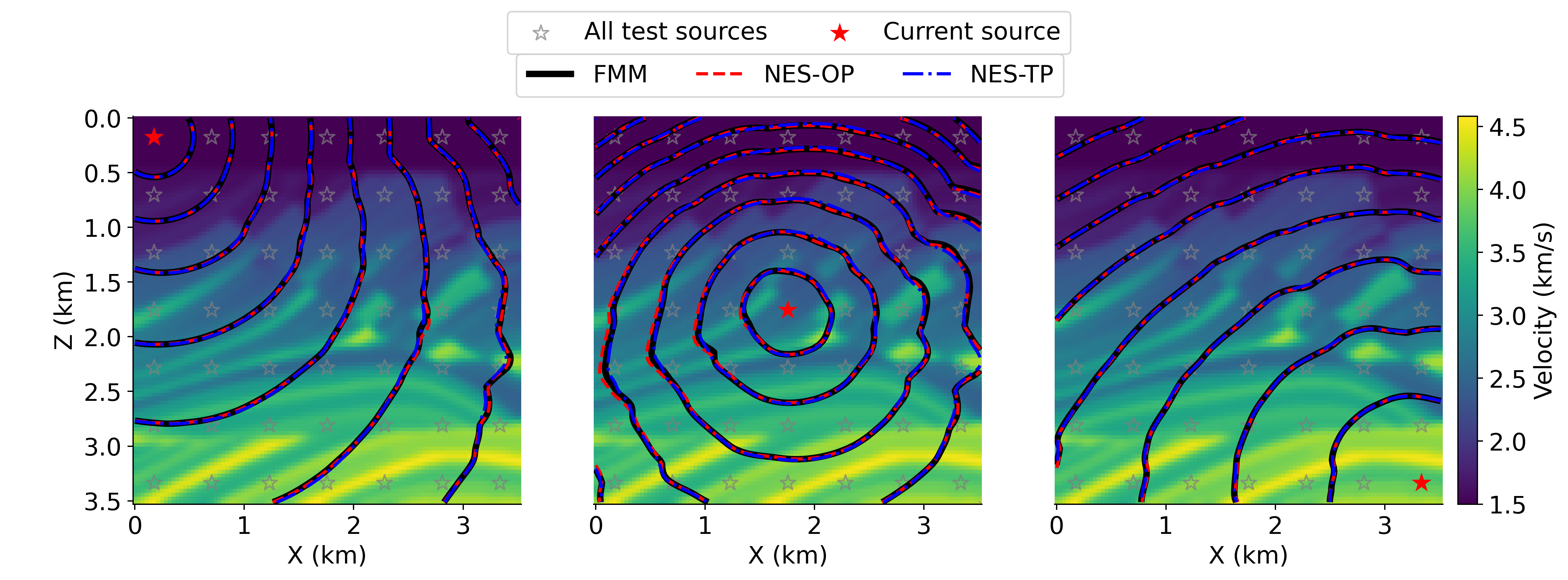}
    \caption{The solutions of NES-OP and NES-TP comparatively to the reference solution. Red star indicates the current source location, the gray star show all considered source locations.}
    \label{fig:NES_Marmousi}
\end{figure*}

Here we provide the test of the NES on the most heterogeneous part of the Marmousi model which was smoothed by a gaussian filter. For the test we chose $7\times7$ sources (49 in total) with $101\times101$ receivers to compare with the reference solution. Sources and receivers are evenly spaced in the domain of the velocity model. The sample solutions are presented in Figure~\ref{fig:NES_Marmousi} for the three different sources. Note, that a single source means a single NES-OP, that is, we have 49 different NES-OPs for 49 sources, whereas a single NES-TP is defined for any number of sources. All 49 NES-OPs and the single NES-TP were trained for 3000 epochs. For each NES-OP the training set was $101\times101-1$ (10200), whereas for the NES-TP the training set contained $49\times101\times101-49$ (499800) points. Again, the source points are always excluded from the training set due to the singularity. Overall, the average RMAE over 49 sources was about 0.5$\%$ and 0.8$\%$ for the NES-OP and the NES-TP respectively. 

\begin{figure*}
    \centering
    \includegraphics[width=0.7\textwidth]{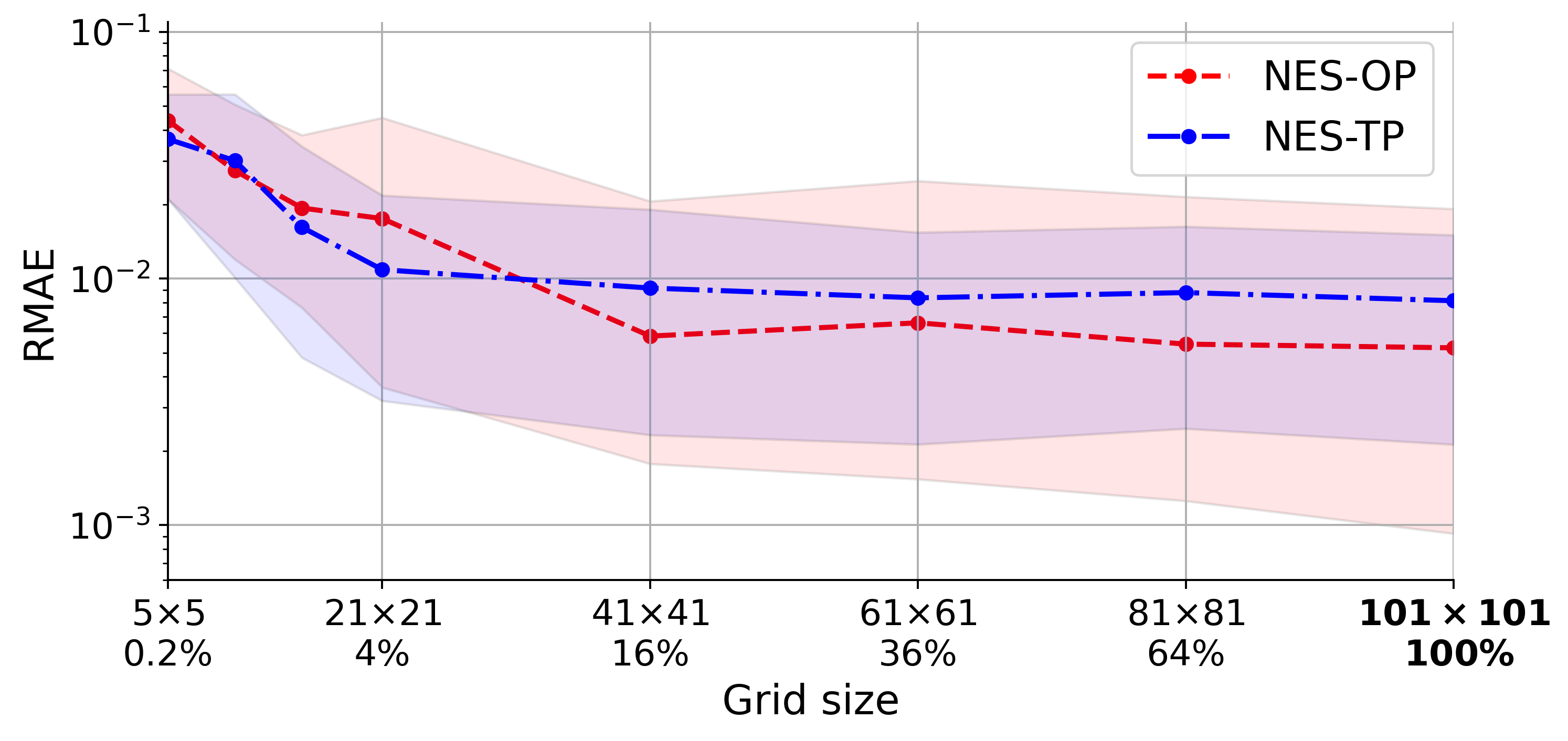}
    \caption{The convergence of the NES-OP and the NES-TP in the Marmousi model. The X-axis indicates the grid size of the receiver points for each source location and the percentage from the original grid size highlighted by bold font, and on which the RMAE was measured. The lines show the average RMAE over all 49 sources, the transparent areas indicate the min-max band over 49 sources.}
    \label{fig:NES_Marmousi_converge}
\end{figure*}

In addition, we tested the convergence of the NES-OP and the NES-TP depending on the sparsity of the receiver points in the training set and the results are shown in Figure~\ref{fig:NES_Marmousi_converge}. From the Figure~\ref{fig:NES_Marmousi_converge} we can conclude that the difference between the NES-OP and the NES-TP is insignificant but systematic. The NES-TP provides slightly higher errors but more stable solution (min-max band is narrower). One can notice that the RMAE for both almost stopped decreasing for the grid sizes bigger than $41\times 41$. That can be due to the fact that the number of the training epochs, the number of trainable weights, and the sampling of collocation points remained fixed throughout all the experiments, while the information about the velocity model was increasing. This problem of weak convergence is known for PINNs and can be partially resolved using proper techniques for the sampling of training points \cite{yu2022gradient}, but we did not consider this aspect in our paper. 

The summary of the performance with the sizes of networks and elapsed time for the training is presented in Table~\ref{tab:MarmousiSummary}. One can notice that the NES-TP is the most attractive as it has the fewest number of trainable weights, require the acceptable training time and provides the RMAE < 1\%. The NES-OP could be used to get a lower error for one specific source. Comparing the NES-TP with the FMM, one can notice that the NES-TP inference time (after training) is comparable with the FMM, and the NES-TP size in memory is significantly lower than the FMM. The FMM size will grow linearly with the increasing grid size, whereas the NES-TP will grow much slower (if we decide to increase number of hidden layers or units). These tests were performed with the fixed complexity of the NES-OP and the NES-TP, but to get higher accuracy more epochs of training and more hidden layers and units in networks can be used. Overall, the NES-TP can be a good alternative to the FMM for the problems with many sources as it gives the compact representation with good accuracy. In addition, although the training time will grow in 3D models, the inference time will not grow as fast as the time of FMM. Our preliminary tests showed that in 3D models, the NES-TP provides 4-5 times faster inference time.

\begin{table}
    \centering
    \caption{The summary of the performance of the NES-OP and the NES-TP in the Marmousi model for the considered 49 sources. The data are provided for the the grid sizes 41$\times$41 and 101$\times$101 in the first and the second rows for the NES-OP and the NES-TP respectively. These grid sizes are chosen as the lower and the upper bound when the RMAE almost stopped decreasing (see Figure~\ref{fig:NES_Marmousi_converge}). These tests were performed on GPU RTX 3070 Laptop. Batch size is always 1/4 of the total number of training points. Inference time column indicates the run-time on a given set of points once a network is trained. Size column indicates the number of the trainable weights for the NES-OP and the NES-TP, and the grid size for the FMM. Note, that the data for the NES-OP contain 49 separate networks.}
    \label{tab:MarmousiSummary}
\begin{tabular}{|c||c|c|c|c|c|}
    \hline
    Solver & RMAE, \% & \makecell{Elapsed time \\ for 3000 epochs, (s)} & Inference time, (s) & Training points & Size \\ \hline \hline
    \multirow{2}{4em}{\textbf{NES-OP}} & 0.6 & 2500 & 0.8 & 82320 & \multirow{2}{4em}{\makecell{852894}} \\ 
    & 0.5 & 3000 & 1 & 499800 & \\ \hline
    \multirow{2}{4em}{\textbf{NES-TP}} & 0.9 & 200 & 0.07 & 82320 & \multirow{2}{4em}{\makecell{17558}} \\ 
    & 0.8 & 960 & 0.3 & 499800 & \\ \hline
    FMM & - & - & 0.2 & - & \makecell{499849} \\ \hline
\end{tabular}
\end{table}

\subsection{Comparison with EikoNet and PINNeik}
In this section, we give a comparison with the existing solutions for solving the eikonal equation: PINNeik \cite{bin2021pinneik} and EikoNet \cite{smith2020eikonet}. We do separate comparison because PINNeik works for the one-point eikonal, but the EikoNet uses two-point eikonal. For these tests, we used the same Marmousi model from Figure~\ref{fig:NES_Marmousi}. In these tests, we used the different number of hidden layers and the training set sizes to show a possible variability in the performance. The results of the comparisons are shown in Table~\ref{tab:NESvsPINNeikvsEikoNet}.

The PINNeik uses the conventional factorization (eq. \ref{eq:FactorizationOP}) and the loss function based on the $L_2$-norm which includes the equation, the positivity and the boundary conditions. The PINNeik architecture consists of 10 hidden layers with 20 units on each, and locally-adaptive arctangent activation \cite{jagtap2020locally}. The NES-OP architecture contained 4 hidden layer with 50 units on each layer. The total number of collocation points for training is $300\times281$ evenly spaced in the domain. The source is positioned at the center of the model as it is shown in central panel in Figure~\ref{fig:NES_Marmousi}. Number of epochs was 5000. We made 5 independent launches for both solvers to see the average result. From the Table~\ref{tab:NESvsPINNeikvsEikoNet} one can notice that the NES-OP is faster and significantly outperforms PINNeik in accuracy (60 times lower RMAE).

The EikoNet also uses the conventional factorization but for the two-point eikonal (eq. \ref{eq:FactorizationTP}). The EikoNet loss function based on $L_2$-norm and is written for the difference between the predicted and true velocities with respect to the receiver-part equation only (based on the first equation in \ref{eq:TwoPointEikonal}) and without the reciprocity. The EikoNet uses deep architecture with ten residual blocks and $elu$ activation. The NES-TP contained 6 hidden layers with 100 units on each. For the training set, we used random 250000 source-receiver pairs. With that, EikoNet utilized random distance sampling, whereas NES-TP utilized random uniform distribution in the domain. We made five independent launches with 1000 epochs for both solvers. The errors were estimated on the three test sources with positions as shown in Figure~\ref{fig:NES_Marmousi}. From the Table~\ref{tab:NESvsPINNeikvsEikoNet} one can notice that the NES-TP provides: 13 times lower error, 32 times faster training, 154 times more compact architecture. 

\begin{table}
    \centering
    \caption{Comparison of NES-OP with PINNeik and NES-TP with EikoNet. RMAE indicates the average for five independent launches. Training time shows the average time: for 5000 epochs of NES-OP and PINNeik; for 1000 epochs of NES-TP and EikoNet. These tests were performed on GPU Tesla P100-PCIE.}
    \label{tab:NESvsPINNeikvsEikoNet}
\begin{tabular}{|c||c|c|c|}
    \hline
    Solver & RMAE, \% & Elapsed time, (s) & Size \\ \hline \hline
    \textbf{NES-OP} & \textbf{0.2} & \textbf{240} & \textbf{7856} \\ \hline
    PINNeik & 12.4 & 330 & 4061 \\ \hline \hline
    \textbf{NES-TP} & \textbf{0.4} & \textbf{300} & \textbf{51308} \\ \hline
    EikoNet & 5.4 & 9600 & 7913249 \\ \hline
\end{tabular}
\end{table}

%% file: Discussion.tex
\section{Discussion} \label{Discussion}
We presented the Neural Eikonal Solver for the one-point and the two-point eikonal equations. We discussed the crucial aspects of the proposed NES architecture: the improved factorization, the loss function, the activation function, the reciprocity principle, the type of a distribution for the random weights initialization, the input data scaling, and the type of connection between the hidden layers. These helped to enhance significantly the performance of the NES. Some aspects still remain unresolved, which we plan to consider as a future development.

\subsection{Future developments}

The sampling of the collocation points for the training is very important for the accuracy and the convergence. We considered only the influence of regular grids on the accuracy (see Figure~\ref{fig:NES_Marmousi_converge}), but thorough investigation is needed to better understand what sparsity of the training set is enough to get the highest performance of the NES. Our preliminary tests showed no visible and systematic improvements for the following techniques: gradients based sampling (more points where gradient of velocity model is high), importance-based sampling \cite{nabian2021efficient}, and weighted sampling \cite{smith2020eikonet}. However, the Residual-based Adaptive Refinement (RAR) \cite{lu2021deepxde} (adding points where needed) showed promising results by providing better accuracy with lower number of collocation points.

The neural-network architecture is important for the performance. Preliminary, we found that the total number of trainable weights between $10^3$-$10^4$ provides good accuracy, but we did not consider how exactly the number of trainable weights increases with the complexity of the velocity model. In our convergence test on the Marmousi model (see Figure~\ref{fig:NES_Marmousi_converge}), we used the fixed architectures which could be a reason why the errors almost stopped decreasing after a certain grid size. Therefore, a deeper investigation is needed to understand how the neural network depth should grow with the increasing information about the velocity model heterogeneities. 

It has been shown that the pre-trained neural network in one velocity model can be used for another model to provide a faster convergence \cite{smith2020eikonet, bin2021pinneik}. The transfer learning can be beneficial for reducing the training time, but we did not consider this problem in our work.

In our paper we mitigated the challenges related to the caustics using a non-symmetric loss function with the Hamiltonian form of the eikonal with $p=2$. But, usually these challenges are tackled using the viscosity solution which provides the stability and uniqueness even for non-differentiable solutions \cite{barles2013introduction}. The trick is to use the additional viscosity term in the eikonal equation $-\varepsilon\Delta\tau$ (laplacian) by vanishing the degree of viscosity $\varepsilon \rightarrow 0$. The numerical algorithms, such as the FMM, do not solve the extended equation with viscosity term, but utilize the discrete approximation of the viscosity solution \cite{monneau2010introduction}. We tested the influence of the viscosity term $-\varepsilon\Delta\tau$ on the NES training dynamic, but it only decreased the accuracy. The future investigations of the sophisticated neural-network architectures representing the viscosity solution \cite{darbon2021some} may be helpful for the eikonal equation.



\subsection{Potential applications}
The scope of our work focused on the isotropic eikonal equation only, but the NES can be extended to the various anisotropic eikonal equations in a similar way as it was shown for the PINNeik \cite{bin2021pinneik}. As well, the massive computations of traveltimes for large number of source-receiver pairs (look-up tables) require a lot of memory for storage, whereas the NES-TP solution has the fixed size defined by the complexity of the neural-network architecture thereby providing drastic compression opportunities. The compression of the look-up tables becomes very useful in the Kirchhoff migration procedures \cite{alkhalifah2011efficient}, which may require large tables (several Tb), while the neural network takes hundreds Kb, which was shown on the example of the approximation of the eikonal solution \cite{grubas2020traveltime, grubas2020seismic}. Other possible applications for the NES-TP: traveltime simulation of the reflected waves according to the Huygens principle; the ray multipathing analysis \cite{rawlinson2010multipathing, smith2020eikonet}; the earthquake localization \cite{smith2021hyposvi, grubas2021localization}; the traveltime tomography \cite{silvaphysics, waheed2021pinntomo}.







%% file: Conlusions.tex
\section{Conclusions} \label{Conclusions}

In this paper, we introduced the Neural Eikonal Solver framework for solving the one-point and the two-point eikonal equation using the PINN concept. We suggested several novelties to the neural-network architecture to improve its accuracy and to speed up training. We proposed the improved factorization of the eikonal equation to speed up the training by constraining the NES between the fastest and the slowest solutions. To address the intrinsic pathologies related to the caustic singularities, the NES involves a non-symmetric loss function based on the $L_1$-norm and a Hamiltonian form of the eikonal equation, and uses the gaussian activation function. To account for the reciprocity principle we proposed a symmetrization that provides a strict observance of the reciprocity. We also suggested the random weight initialization based on $he\_normal$ distribution and the scaling of the input data by a maximum value. Almost all of the proposed features require no additional computational cost. This resulted in the improvement of the accuracy of traveltime computations by the two orders of magnitude.

The benchmark testing conducted on a simple model with caustics showed that the NES reaches the relative mean absolute error of about 0.07-0.16\% from the second-order factored FMM. For complicated Marmousi model the NES reaches error $<0.5\%$ after several minutes of training. The test showed that the NES significantly outperforms existing PINN solutions to the eikonal equation showing 10-60 times smaller error and 2-30 times faster training. The NES-OP provides higher accuracy than the NES-TP for one specific source. The NES-TP provides slightly lower accuracy than the NES-OP but gives a compact representation of traveltimes for arbitrary source-receiver pairs which is perspective for applications in seismic exploration. In this paper, we showed 2D examples for the simplicity but extension to 3D is straightforward with the accuracy and the performance remaining the same. 